\documentclass{article}

\usepackage{microtype}
\usepackage{graphicx}
\usepackage{subcaption}
\usepackage{booktabs} 

\usepackage{hyperref}

\usepackage[preprint]{icml2026}

\usepackage{amsmath}
\usepackage{amssymb}
\usepackage{mathtools}
\usepackage{amsthm}

\usepackage[capitalize,noabbrev]{cleveref}

\theoremstyle{plain}

\theoremstyle{definition}

\theoremstyle{remark}

\usepackage{multirow}

\usepackage[textsize=tiny]{todonotes}

\icmltitlerunning{Tutti: Expressive Multi-Singer Synthesis via Structure-Level Timbre Control and Vocal Texture Modeling}

\begin{document}

\twocolumn[
  \icmltitle{Tutti: Expressive Multi-Singer Synthesis via Structure-Level Timbre Control \\ and Vocal Texture Modeling}



  \icmlsetsymbol{equal}{*}

  \begin{icmlauthorlist}
    \icmlauthor{Jiatao Chen}{whut,wechat}
    \icmlauthor{Xing Tang}{whut}
    \icmlauthor{Xiaoyue Duan}{wechat}
    \icmlauthor{Yutang Feng}{wechat}
    \icmlauthor{Jinchao Zhang}{wechat}
    \icmlauthor{Jie Zhou}{wechat}
  \end{icmlauthorlist}

  \icmlaffiliation{whut}{Wuhan University of Technology, School of Computer Science and Artificial Intelligence, Wuhan, China}
  \icmlaffiliation{wechat}{WeChat AI, Tentenct Inc, China}

  \icmlcorrespondingauthor{Xing Tang}{tangxing@whut.edu.cn}
  \icmlcorrespondingauthor{Jinchao Zhang}{dayerzhang@tencent.com}

  \icmlkeywords{}

  \vskip 0.3in
]



\printAffiliationsAndNotice{}  

\begin{abstract}
    While existing Singing Voice Synthesis systems achieve high-fidelity solo performances, they are constrained by global timbre control, failing to address dynamic multi-singer arrangement and vocal texture within a single song. To address this, we propose \textbf{Tutti}, a unified framework designed for structured multi-singer generation. Specifically, we introduce a Structure-Aware Singer Prompt to enable flexible singer scheduling evolving with musical structure, and propose Complementary Texture Learning via Condition-Guided VAE to capture implicit acoustic textures (\textit{e.g.}, spatial reverberation and spectral fusion) that are complementary to explicit controls. Experiments demonstrate that Tutti excels in precise multi-singer scheduling and significantly enhances the acoustic realism of choral generation, offering a novel paradigm for complex multi-singer arrangement. Audio samples are available at \url{https://annoauth123-ctrl.github.io/Tutii_Demo/}.
\end{abstract}

\section{Introduction}

Music is a universal language for human emotional expression. With the rapid development of deep learning, Generative AI has made remarkable breakthroughs in music generation \cite{dhariwal2020jukebox,agostinelli2023musiclm,liu2023audioldm}, demonstrating the potential to create complete musical works.

Within musical compositions, the singing voice stands out as the most infectious and expressive component, carrying profound narrative power. Consequently, Singing Voice Synthesis (SVS), as a unique domain combining linguistic content with musical expression, has achieved similarly significant accomplishments. Leading models like DiffSinger \cite{liu2022diffsinger} and VISinger \cite{zhang2022visinger} have established high-fidelity generation paradigms, enabling the synthesis of realistic solo performances.

However, despite the excellence of existing SVS models in timbre cloning, they remain largely constrained by the ``Soloist Paradigm''. The vast majority of existing models assume a single-singer input for a single song or only support stable singer identity, where the singer identity is defined as a time-invariant global condition that remains constant throughout the entire piece. This rigid setting ignores two dimensions crucial to complex musical expression:
(1) \textit{Singer Arrangement}—the temporal scheduling of multiple singers according to musical structure. In real compositions, singers' roles shift with the song's progression: a verse may feature a solo vocalist, while the chorus brings in additional singers or a full choir. Furthermore, transitions frequently occur between different soloists (\textit{e.g.}, solo-to-solo) across sections. Existing models fundamentally lack the mechanism to model such singer transitions and the simultaneous timbre representation required for ensembles within a single song.
(2) \textit{Vocal Texture}—the emergent acoustic properties arising from vocal interactions and production, including spatial reverberation, spectral fusion, and harmonic layering. Multi-person choruses are not simple linear mixtures of individual voices; they involve complex interactions such as the blending of formants across different vocal parts. Current SVS models, designed primarily for solo synthesis, have no explicit mechanism to model these interactions, resulting in generated multi-person segments that sound like artificially mixed individual tracks rather than cohesive ensemble performances.

\begin{figure*}[tb!]
	\centering
	\includegraphics[width=1.9\columnwidth]{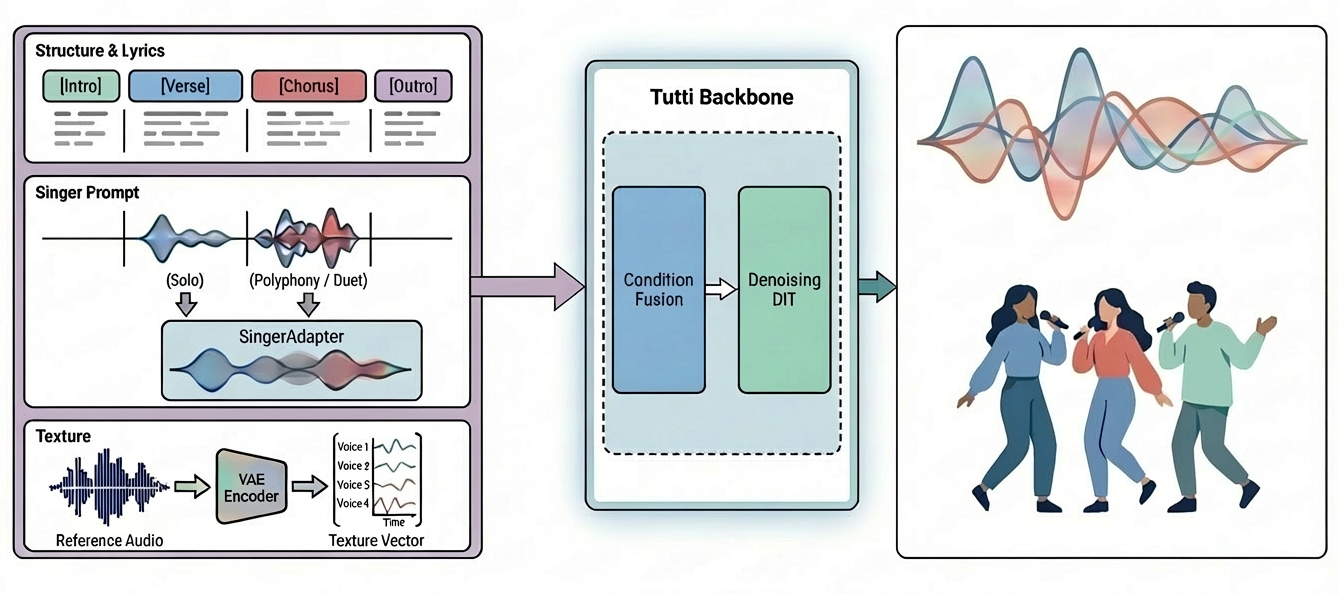}
	\caption{The overview of the Tutti framework for structure-aware multi-singer generation. The workflow begins by constructing structure-aware singer prompts and extracting complementary texture features from reference audio. These conditions, along with lyrics and structure labels, are fed into the DiT backbone to generate target vocal latents. Finally, the Vocal VAE decoder reconstructs the latents into high-fidelity multi-singer waveforms.}
	\label{fig:framework}
\end{figure*}

To fill this gap, we extend the task from traditional SVS to Multi-Singer Singing Voice Synthesis, and propose \textbf{Tutti}: a structure-aware framework designed for complex multi-singer arrangement. Specifically, we design a Structure-Aware Singer Prompt, capable of constructing singer sequences based on musical structure (\textit{e.g.}, verse/chorus) to flexibly schedule the participation of multiple singers. To address the challenge of representing multiple singers simultaneously, we further incorporate an Adaptive Singer Prompt Fuser to dynamically integrate multi-singer embeddings. Simultaneously, to capture the rich acoustic nuances beyond explicit features, we propose Complementary Texture Learning via Condition-Guided VAE. This module learns to extract implicit texture features—such as mixing effects and spatial atmosphere—that are complementary to explicit conditions like lyrics and singer identity, thereby precisely restoring the acoustic quality of vocal interactions. Our contributions can be summarized as follows:

\begin{itemize}
\setlength{\itemsep}{0pt}
    \item We propose Tutti, the first generative framework designed for multi-singer synthesis within a single song.
    \item We design a Structure-Aware Singer Prompt combined with an Adaptive Singer Prompt Fuser, which flexibly handles complex transitions between soloists and ensembles, enabling precise control over the singer composition of each section.
    \item We propose Complementary Texture Learning via Condition-Guided VAE, which introduces a condition-guided reconstruction task to capture implicit vocal textures complementary to explicit controls, significantly enhancing the acoustic realism of the generation.
\end{itemize}

\section{Related Work}

\subsection{Singing Voice Synthesis}

In recent years, the field of Singing Voice Synthesis (SVS) has undergone a paradigm shift from end-to-end generation to diffusion models, achieving significant progress in sound quality, control granularity, and input flexibility. The VISinger \cite{zhang2022visinger} series and DiffSinger \cite{liu2022diffsinger} established a high-fidelity generation paradigm based on explicit score input, with DiffSinger successfully introducing diffusion models to SVS, significantly improving synthesis naturalness and stability. Subsequently, researchers have dedicated efforts to expanding the task boundaries and control capabilities of SVS: MVoice \cite{huang2023make} proposed a unified framework integrating TTS, VC, and SVS, preliminarily achieving zero-shot multi-singer adaptation via Speaker Prompts; TCSinger \cite{zhang2024tcsinger} and ExpressiveSinger \cite{dai2024expressivesinger} focused on cross-lingual style transfer and fine-grained performance control (\textit{e.g.}, vibrato, loudness), enhancing generated musical expressiveness. In terms of generation efficiency, InstructSing \cite{zeng2024instructsing} combined DDSP with adversarial training to drastically reduce the computational cost of high-sampling-rate synthesis.

Recent work has further broken the dependence of traditional SVS on musical scores. TechSinger \cite{guo2025techsinger} introduced a flow matching framework combined with a technique predictor, achieving fine-grained modeling of multiple languages and techniques; Vevo \cite{vevo} and Vevo2 \cite{zhang2025vevo2} realized zero-shot timbre imitation and score-free generation through self-supervised disentanglement and multi-modal input mechanisms, driving SVS towards a more flexible generation paradigm. However, while these models excel in single-singer timbre cloning and style control, they remain primarily confined to ``single-song single-singer'' or ``global timbre control'' frameworks, failing to effectively address the complex problem of ``dynamic multi-singer arrangement within a single song,'' and lacking explicit modeling of part interaction and vocal texture in multi-person choruses.

\subsection{Multi-Talker Conversational Generation}

To overcome the limitation of traditional SVS supporting only single-person singing and to realize dynamic multi-singer arrangement within a single song, we draw inspiration from the field of Multi-Talker Conversational Generation. This field is dedicated to solving ``Speaker Scheduling'' and ``Turn-taking'' problems, providing an important reference paradigm for our dynamic singer prompt design.
Early explorations such as dGSLM \cite{nguyen2023generative} employed a dual-tower Transformer architecture, establishing connections between discrete unit streams via cross-attention mechanisms to implicitly learn speaker interaction rhythms and paralinguistic behaviors. Subsequently, to more precisely control Overlapping Speech, CoVoMix \cite{zhang2024covomix} and CoVoMix2 \cite{zhang2025covomix2} introduced a fully non-autoregressive Flow Matching framework, directly mapping multi-stream transcriptions to mixed mel-spectrograms, achieving fine-grained temporal control and speaker separation.
Recent work has further strengthened the explicit modeling of interaction logic. FD-SLMs \cite{wang2024full} proposed a Neural Finite State Machine (Neural FSM) based on Large Language Model (LLM) prediction, determining turn-taking through ``speak/listen'' state switching, realizing online interaction decisions based on semantic context. DialoSpeech \cite{xie2025dialospeech} combined LLMs with dual-track token generation, using special speaker change tokens (\textit{e.g.}, `[spkchange]') to explicitly control the dialogue flow. Building on this, JoyVoice \cite{yu2025joyvoice} employed explicit speaker ID sequences and embeddings as conditions, achieving structured scheduling of long-context dialogues via a causal autoregressive Transformer, making significant progress in speaker consistency.
However, directly transferring the aforementioned conversational generation paradigms to multi-singer choral tasks faces fundamental challenges. First, dialogue models typically treat overlapping speech as temporal preemption or simple signal mixing, lacking acoustic modeling of vocal texture—the harmonic fusion and resonance of multiple parts in the frequency domain. Second, dialogue scheduling logic is primarily semantically driven, whereas singing arrangement is constrained by strict musical structure (\textit{e.g.}, verse/chorus, harmonic progression). Existing dialogue models cannot perceive this high-level musical prior, resulting in generated choral segments that often lack pitch harmony between parts and overall cohesion. 

\section{Methodology}

\begin{figure*}[tb!]
	\centering
	\includegraphics[width=1.9\columnwidth]{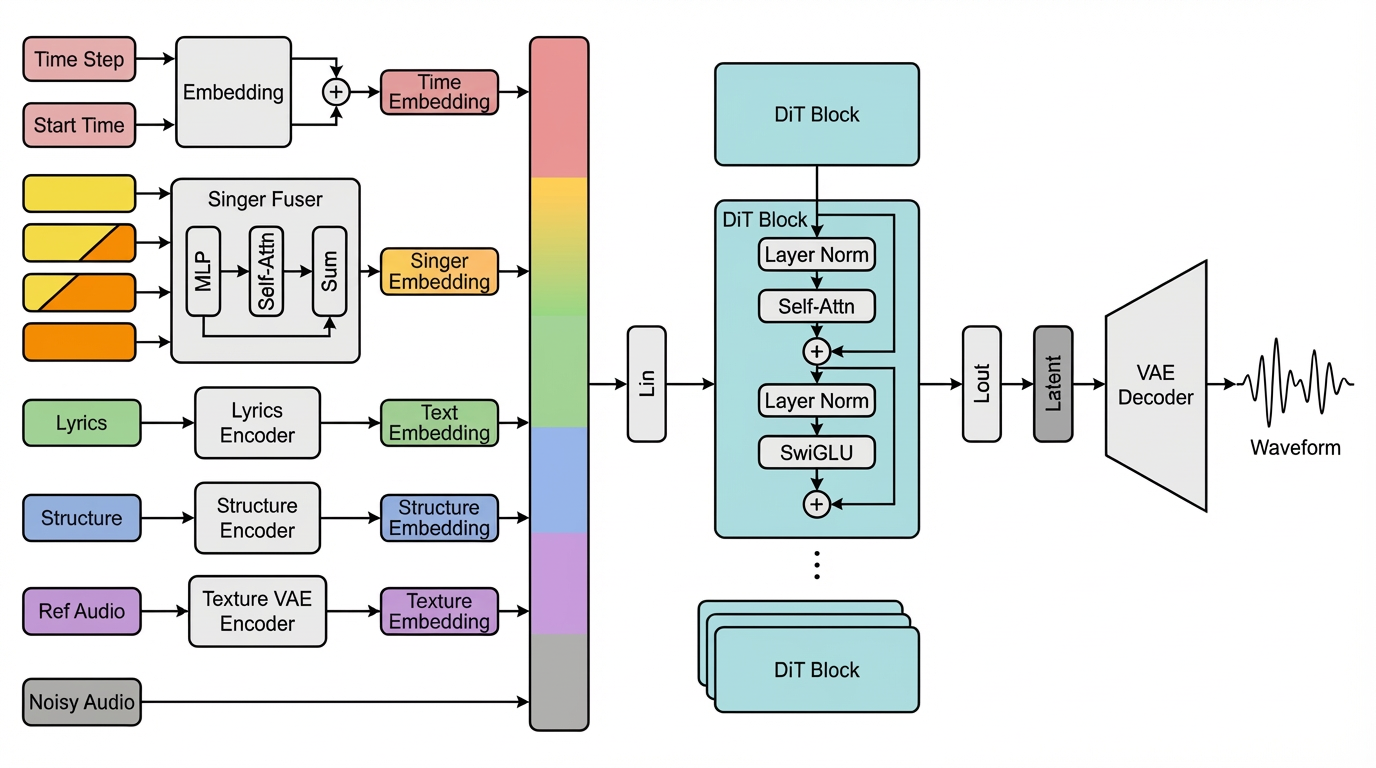}
	\caption{The architecture of the DiT-based backbone. The left panel illustrates the condition processing modules: the structure-aware singer prompt (denoted in orange) utilizes an adaptive fuser to integrate multi-singer information, while the texture embedding (denoted in purple) is extracted from reference audio to provide complementary acoustic features. These embeddings are concatenated with other conditions and input into the DiT blocks. Finally, the predicted latents are decoded by the Vocal VAE decoder to generate the waveform.}
	\label{fig:Architecture}
\end{figure*}

\subsection{Overview}
\label{sec:overview}
We propose \textbf{Tutti}, a structure-aware multi-singer generation framework built upon the Latent Diffusion Transformer (DiT) paradigm. As illustrated in Figure \ref{fig:Architecture}, our model consists of two core components: a Vocal VAE for compressing raw audio into latent representations, and a DiT-based backbone for conditional generation.

\textbf{Model Architecture.}
To model long-sequence musical dependencies efficiently, our backbone adopts a LLaMA-based Transformer architecture, following the design of DiffRhythm \cite{ning2025diffrhythm}. For the latent space, we employ a specialized Vocal VAE based on the Stable Audio 2.0 \cite{evans2025stable}, which is retrained on vocal data to faithfully reconstruct high-fidelity waveforms from low-dimensional latents.

\textbf{Model Conditioning.}
To achieve precise control over content, structure, and timbre, we inject multiple conditions into the generation process. Specifically, the lyrics $L$ are encoded via a pre-trained text encoder to provide content guidance, while the structure label $S$ represents musical sections (\textit{e.g.}, verse, chorus) to guide the structural progression. To handle complex singer scheduling, we employ a structure-aware singer prompt $C_{singer}$ as a dynamic representation of multi-singer identities, which is processed by our proposed Adaptive Singer Prompt Fuser. Furthermore, we extract vocal texture $Z_{texture}$ from reference audio via our Condition-Guided VAE encoder to capture implicit acoustic details such as breathiness and spatial effects. Finally, we incorporate both the diffusion time step $t$ and the audio start time to inform the model of the denoising progress and the absolute temporal position within the song.

Based on these conditions, we formulate the generation task as learning a conditional distribution $p_\theta(z_0 | L, S, C_{singer}, Z_{texture})$, where $\theta$ represents the learnable parameters.

\textbf{Features Fusion.}
We employ a concatenation-based fusion strategy. Continuous conditions (\textit{e.g.}, singer prompts, texture) and discrete conditions (\textit{e.g.}, lyrics, structure) are aligned to the latent frame rate via broadcasting or downsampling, and then concatenated with the noisy latent $z_t$ along the feature dimension. This ensures the DiT can attend to all control signals simultaneously at every step.

\textbf{Training Objectives.}
For the Vocal VAE, we employ a composite objective consisting of multi-resolution STFT loss \cite{steinmetz2020auraloss} for spectral convergence, KL divergence loss \cite{yu2013kl} for latent regularization, and adversarial loss to enhance perceptual quality. Subsequently, for the DiT backbone, following the conditional flow matching paradigm~\cite{yaron2023flowmatching}, our model learns a velocity field $v_\theta(z_t, t, c)$ that transports the noise distribution $p_0(z)$ to the data distribution $p_1(z)$ through the ODE:
\begin{equation}
    \frac{dz_t}{dt} = v_\theta(z_t, t, c) \quad \text{with} \quad 
    \begin{cases}
        z_0 \sim p_0(z) \\
        z_1 \sim p_1(z)
    \end{cases}
\end{equation}
The training objective minimizes the expected squared error between predicted and target velocity fields:
{\small
\begin{equation}
    \mathcal{L} = \mathbb{E}_{t \sim \pi_{\text{ln}}, z_t \sim p_t(z_t)} \left[\|v_\theta(z_t, t, c) - (z_1 - z_0)\|_2^2\right],
\end{equation}
}where $c$ denotes the set of conditions, and the timestep sampling distribution $\pi_{\text{ln}}(t;m,s)$ follows the logit-normal density:
{\small
\begin{equation}
    \pi_{\text{ln}}(t;m,s) = \frac{1}{s\sqrt{2\pi}}\frac{1}{t(1-t)} \exp\left(-\frac{(\mathrm{logit}(t)-m)^2}{2s^2}\right),
\end{equation}
}with $\mathrm{logit}(t) = \log\frac{t}{1-t}$. As discussed in Stable Diffusion 3~\cite{esser2024scalingrectifiedflowtransformers}, logit-normal sampling provides adaptive weighting where the scale parameter $s$ controls concentration around mid-point timesteps, while the location parameter $m$ enables bias toward either data or noise domains. In practice, we sample $u \sim \mathcal{N}(m, s)$ and map it through the logistic function $t = \sigma(u) = 1/(1+e^{-u})$.

\subsection{Structure-Aware Singer Prompt}
\label{sec:structure-aware singer prompt}
To enable flexible arrangement of multiple singers within a single song, we design a structure-aware prompt construction strategy that automatically schedules singer resources based on musical structure, followed by an adaptive fuser to integrate multi-singer information.

\textbf{Prompt Construction.}
We propose a structure-guided pipeline to construct the singer prompt. First, we utilize SongPrep~\cite{tan2025songpreppreprocessingframeworkendtoend} to segment the input audio timeline $T$ into $M$ non-overlapping structural units $\{\tau_1, \tau_2, \dots, \tau_M\}$. 1) For each segment $\tau_m$, we employ CAM++~\cite{wang2023camfastefficientnetwork} to extract a representative singer embedding $\mathbf{v}_m$. 2) Based on the assumption that verse sections typically feature a single vocalist, we compute the cosine similarity $s_{i,j}$ solely between embeddings of verse segments. Pairs satisfying $s_{i,j} < \delta_{id}$ are considered as different singers, forming the global singer set $U_{global} = \{\mathbf{u}_1, \dots, \mathbf{u}_K\}$ for the song. 3) We perform structure-based assignment: for verse segments, we strictly assign the unified singer embedding from the identified cluster to enhance zero-shot timbre stability; for chorus or bridge segments, if the global singer count $K > 1$, we calculate the similarity between the current segment's embedding $\mathbf{v}_m$ and each singer $\mathbf{u}_k$ in $U_{global}$ as $s_{m,k}$. When the count of matched singers satisfies $\sum_{k=1}^K \mathbb{I}(s_{m,k} > \delta_{multi}) \ge 2$, where $\mathbb{I}(\cdot)$ is the indicator function, the segment is identified as a multi-singer section, and we simultaneously update its structure label to reflect this state.

\textbf{Adaptive Singer Prompt Fuser.}
To effectively fuse the identified singer embeddings for each segment into a unified representation, we design an adaptive fuser based on the self-attention mechanism. Given the singer set $U_m$ for the $m$-th segment, the fuser maps them to a single condition embedding $C_{singer}$. Specifically, the generated condition embedding is calculated as:
\begin{equation}
C_{singer} = \sum_{i=1}^{k_m} \text{Softmax}\left(\frac{Q K^T}{\sqrt{d_k}}\right)_i \mathbf{u}_i
\end{equation}
Here, the query ($Q$), key ($K$), and value ($V$) are all linearly projected from the input singer embeddings $U_m$. This parameterized weighting scheme allows the model to adaptively handle the primary-secondary relationships between lead vocals and harmonies. The resulting embedding is then broadcast to all time steps within the segment, providing a stable yet structure-aware timbre control signal.

\subsection{Complementary Texture Learning via Condition-Guided VAE}

\begin{figure}[tb!]
	\centering
	\includegraphics[width=1\columnwidth]{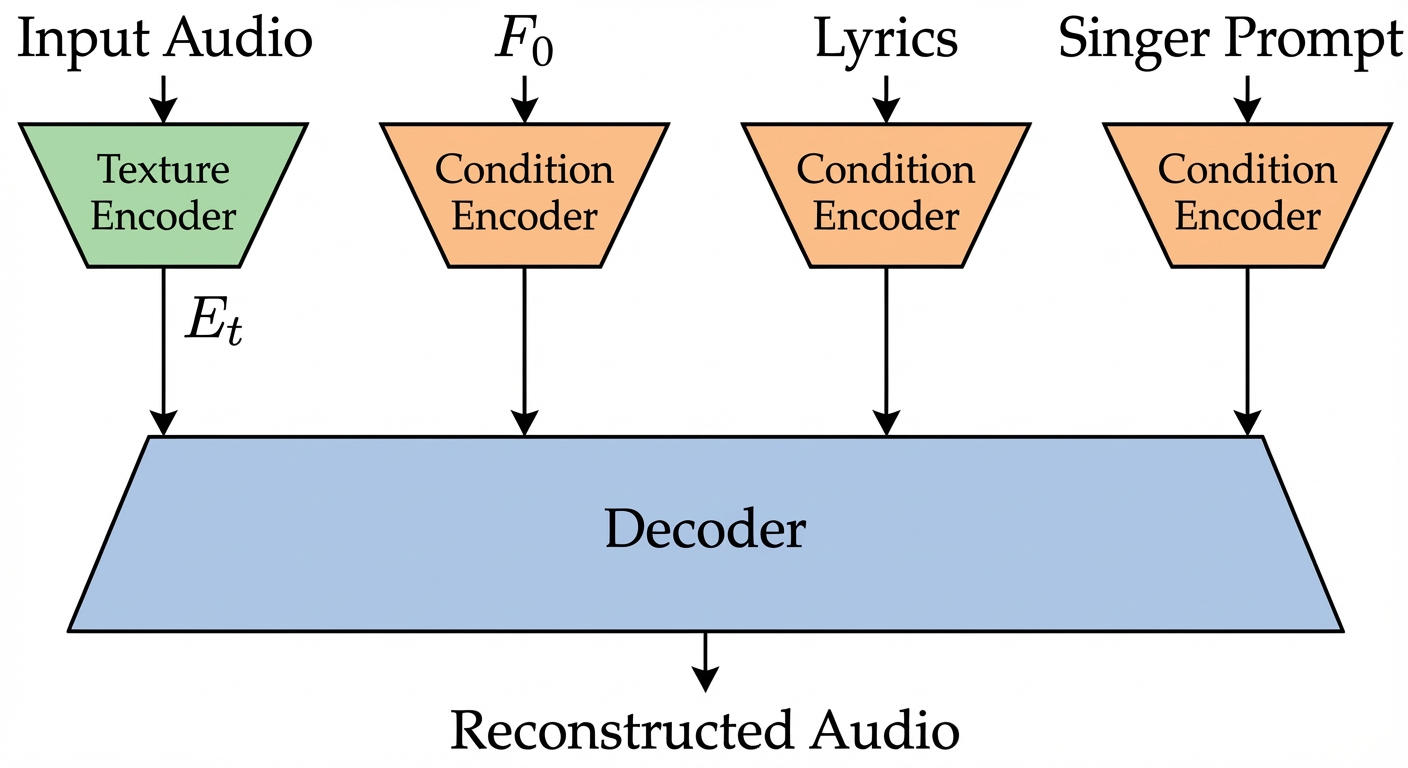}
	\caption{The architecture of Condition-Guided VAE for complementary texture learning.}
	\label{fig:TextureVAE}
\end{figure}

In complex multi-singer scenarios, relying solely on explicit conditions (\textit{e.g.}, lyrics, structure labels, and singer prompts) is often insufficient to generate high-fidelity audio with realistic acoustic nuances. These explicit controls define the content and identity but fail to capture the ``vocal texture''—acoustic properties such as spatial reverberation, mixing effects, and high-frequency harmonics that determine the acoustic realism of the performance. To address this, we introduce a Condition-Guided VAE to learn a complementary texture representation that fills the gap between explicit controls and real audio.

First, we train a dedicated Vocal VAE. Since standard music VAEs \cite{evans2025stable} are optimized for general music and may not capture fine-grained vocal characteristics, we retrain the VAE architecture described in Section \ref{sec:overview} specifically on our vocal dataset. To encourage the learned latent space to capture texture information complementary to our explicit conditions, we propose a condition-guided reconstruction task. As illustrated in Figure \ref{fig:TextureVAE}, we formulate the VAE training as a latent-perturbation reconstruction problem. Given a clean vocal waveform $y$, the encoder $\mathcal{E}$ first maps it to a latent representation $z$. To prevent the encoder from retaining explicit information (such as pitch, content and timbre) that should be provided by other conditions, we apply gaussian noise perturbation to $z$ to obtain a corrupted latent $\tilde{z}$. Crucially, to disentangle texture from content and timbre, we explicitly provide the decoder $\mathcal{D}$ with the set of explicit conditions $C_{ex} = \{c_{lyrics}, c_{singer}, c_{F0}\}$, where $c_{lyrics}$, $c_{singer}$, and $c_{F0}$ correspond to the embeddings of lyrics, structure-level singer prompts, and fundamental frequency, respectively. The decoder is then tasked with reconstructing the original clean audio $y$ using both the perturbed latent and the explicit conditions:
\begin{equation}
    \hat{y} = \mathcal{D}(\tilde{z}, C_{ex}) = \mathcal{D}(\mathcal{E}(y) + \epsilon, c_{lyrics}, c_{singer}, c_{F0})
\end{equation}
where $\epsilon$ is sampled from a gaussian distribution $\mathcal{N}(0, \sigma^2I)$. By supplying $C_{ex}$ directly to the decoder and perturbing the latent space, we encourage the encoder to discard information already present in these conditions which are robustly provided by $C_{ex}$ and focus on extracting the residual acoustic details—the vocal texture—required to restore the lossless target $y$.

During inference, this design allows for flexible texture control. We can optionally provide a reference audio to the encoder to extract a texture latent $z_{texture}$. Since the encoder is trained to disentangle texture from pitch, content and timbre, this latent primarily captures acoustic properties—such as mixing effects, spatial atmosphere, and timbre stability. This ensures that the reference audio enhances the acoustic realism and richness of the vocal texture, while the musical structure is predominantly controlled by the DiT backbone and explicit conditions.

\begin{table*}[t]
\caption{Comparative evaluation of vocal waveform reconstruction performance. We report STOI, WB-PESQ, MCD, and FAD for both lossless-to-lossless and lossy-to-lossless reconstruction scenarios.}
\label{tab:vae}
\begin{center}
\resizebox{1.9\columnwidth}{!}{
\begin{tabular}{l|cccc|cccc}
\toprule
\multirow{2}{*}{Model} & \multicolumn{4}{c|}{Lossless $\rightarrow$ Lossless} & \multicolumn{4}{c}{Lossy $\rightarrow$ Lossless} \\
\cmidrule(lr){2-5} \cmidrule(lr){6-9}
 & STOI$\uparrow$ & WB-PESQ$\uparrow$ & MCD$\downarrow$ & FAD$\downarrow$ & STOI$\uparrow$ & WB-PESQ$\uparrow$ & MCD$\downarrow$ & FAD$\downarrow$ \\
\midrule
DiffRhythm VAE & 0.7557 & 2.7461 & 2.9580 & 0.8252 & 0.7551 & 2.6934 & 3.0601 & 1.4162 \\
LeVo VAE & 0.7246 & 2.5426 & 3.3435 & 1.2382 & 0.7211 & 2.4792 & 3.2666 & 1.8931 \\
SongBloom VAE & 0.7232 & 2.5384 & 3.6137 & 1.1872 & 0.7195 & 2.4757 & 3.6721 & 1.8239 \\
\midrule
Ours (Vocal VAE) & \textbf{0.8175} & \textbf{3.1602} & \textbf{2.4696} & \textbf{0.5676} & \textbf{0.8163} & \textbf{3.0805} & \textbf{2.4458} & \textbf{1.1697} \\
\bottomrule
\end{tabular}}
\end{center}
\end{table*}




\section{Experiment}

\subsection{Experimental Setup}

\textbf{Data Preparation.}
To train our framework, we constructed two independent datasets for the Vocal VAE and the DiT backbone, respectively.
For the Vocal VAE, we collected approximately 140,000 songs. To obtain clean training targets, we uniformly employed BS-Roformer \cite{lu2024music} to extract vocal tracks from the raw audio. Specifically, to ensure stability during reconstruction, we applied loudness normalization to all extracted vocal segments to unify the dynamic range.
For the DiT Backbone, we expanded the scale to approximately 300,000 songs, primarily sourced from AudioBox \cite{vyas2023audiobox} and SongEval \cite{yao2025songeval}. We followed the same vocal extraction pipeline but introduced a more complex annotation workflow. Specifically, we used WhisperX \cite{bain2023whisperx} for lyric transcription and forced alignment, filtering samples based on confidence scores and retaining only Chinese and English songs. To construct multi-singer sequences, we utilized CAM++ \cite{wang2023camfastefficientnetwork} to extract voice print embeddings for each segment and set a strict cosine similarity threshold as 0.4 to distinguish between singers, ensuring the dataset contains significant timbre variations and singer transitions.

\textbf{Baselines.}
To verify the effectiveness of the proposed framework, we designed comparative experiments involving external baselines and internal variants. For acoustic reconstruction, we select DiffRhythm VAE \cite{ning2025diffrhythm}, LeVo VAE \cite{lei2025levo}, and SongBloom VAE \cite{yang2025songbloom} as baselines, representing state-of-the-art audio compression models in music and singing voice domains. For singing voice generation, given the lack of off-the-shelf multi-singer generation models and considering that our task does not require specific pitch input, we selected Vevo2 \cite{zhang2025vevo2} as a strong baseline, representing the state-of-the-art in multi-modal generation and zero-shot timbre imitation. For internal baselines, we included different variants of Tutti for comparison: (1) Tutti-w/o-Texture: removing the complementary texture learning module and relying solely on basic VAE reconstruction; (2) Tutti-w/o-Fuser: removing the Adaptive Singer Prompt Fuser and using simple linear summation to fuse singer features. This setup aims to simultaneously evaluate the model's advantages over existing SOTA and the contributions of its core components.

\textbf{Evaluation Data.} 
To evaluate model performance across different tasks, we constructed three test subsets. For vocal reconstruction, we randomly selected 30 songs with a sampling rate of 44.1kHz. The vocal tracks were extracted via Music Source Separation and served as reconstruction targets for evaluating different VAE architectures. For the evaluation of basic generation metrics, we randomly sampled 50 songs containing vocals from the dataset. For the evaluation of multi-singer related metrics, we first filtered a batch of candidate samples suspected to contain multiple singers through the voice print similarity threshold in the data annotation pipeline, followed by manual verification to exclude samples that were misidentified due to significant timbre variations within single-person performances. This process resulted in a final set of 20 songs as the multi-singer test set. This subset encompasses various typical multi-singer scenarios including band choruses and male-female duets, effectively examining the model's capabilities in singer scheduling and choral modeling.

\textbf{Evaluation Metrics.}
To comprehensively evaluate generation quality, we adopt an evaluation system combining objective and subjective metrics. For acoustic reconstruction, we employ STOI \cite{Taal2010stoi} to measure intelligibility preservation, WB-PESQ \cite{Rix2001pesq} for perceptual quality assessment, MCD \cite{kubichek1993mcd} for spectral distortion, and FAD \cite{kilgour2019fad} for distribution-level fidelity. For singing voice generation, we use Word Error Rate to evaluate lyric intelligibility \cite{radford2023robust}. Note that since our task is defined as score-free generation where the melody is freely determined by the model, standard pitch accuracy metrics (\textit{e.g.}, F0 RMSE) are not applicable. We also calculate Speaker Similarity (SIM) between generated and reference audio. Considering the multi-singer scenario, SIM is computed as a weighted average based on singer duration to measure identity consistency. For subjective evaluation, we employ MOS-Q and MOS-N to assess Audio Quality and Naturalness, respectively. Furthermore, specifically for the multi-singer task in this paper, we introduce two additional metrics rated on a 1-5 scale: (1) Multi-Singer MOS (MS-MOS): evaluating the fusion of multi-singer timbres, the distinctiveness of each singer, and their stability; (2) Melody MOS (Mel-MOS): specifically assessing the melody and pitch accuracy of the performance.

\subsection{Main Results}

\begin{table*}[t]
\centering
\renewcommand\arraystretch{1.2}
\caption{Objective and subjective evaluation results. w/o Texture removes the complementary texture learning module, and w/o Fuser replaces the Adaptive Singer Prompt Fuser with linear summation.}
\label{tab:main_results}
\resizebox{1.8\columnwidth}{!}{
\begin{tabular}{lcccccc}
\toprule
\multirow{2}{*}{\textbf{Model}} & \multicolumn{2}{c}{\textbf{Objective Metrics}} & \multicolumn{4}{c}{\textbf{Subjective Metrics (95\% CI)}} \\
\cmidrule(lr){2-3} \cmidrule(lr){4-7}
                       & WER$\downarrow$ & SIM$\uparrow$ & MOS-Q$\uparrow$ & MOS-N$\uparrow$ & MS-MOS$\uparrow$ & Mel-MOS$\uparrow$ \\
\midrule
GT (Ground Truth)      & 12.45\%         & -             & 4.50\scriptsize$\pm$0.05   & 4.65\scriptsize$\pm$0.05   & 4.30\scriptsize$\pm$0.05    & 4.16\scriptsize$\pm$0.05 \\
\midrule
Vevo2                  & 16.80\%         & 0.657         & 3.85\scriptsize$\pm$0.12   & 4.01\scriptsize$\pm$0.12   & -                           & - \\
\midrule
Tutti (Ours)           & \textbf{13.50\%} & 0.691        & \textbf{4.12\scriptsize$\pm$0.06} & \textbf{4.12\scriptsize$\pm$0.06} & \textbf{4.02\scriptsize$\pm$0.10} & \textbf{3.89\scriptsize$\pm$0.05} \\
\hspace{1em}w/o Texture & 13.85\%         & \textbf{0.705} & 3.99\scriptsize$\pm$0.08   & 3.97\scriptsize$\pm$0.07   & 3.87\scriptsize$\pm$0.06    & 3.57\scriptsize$\pm$0.12 \\
\hspace{1em}w/o Fuser & 17.25\%         & 0.649         & 4.02\scriptsize$\pm$0.07   & 4.10\scriptsize$\pm$0.07   & 3.61\scriptsize$\pm$0.12    & 3.87\scriptsize$\pm$0.06 \\
\bottomrule
\end{tabular}}
\end{table*}


\textbf{Acoustic Reconstruction.} We first evaluate the waveform reconstruction performance in pure vocal scenarios, comparing our retrained Vocal VAE against three representative baselines. The evaluation protocol encompasses two experimental settings: lossless-to-lossless reconstruction, using lossless audio as input; and lossy-to-lossless reconstruction, using MP3-compressed lossy audio as input while maintaining lossless audio as the reference target. As shown in Table \ref{tab:vae}, our Vocal VAE achieves optimal performance across all metrics under both conditions. In terms of intelligibility preservation, STOI improves by approximately 8\% over the second-best baseline. For perceptual quality and spectral reconstruction, both WB-PESQ and MCD show significant improvements, indicating that the reconstructed audio is perceptually closer to the original signal. Regarding distribution-level fidelity, FAD decreases substantially, suggesting that the overall distribution of reconstructed audio more closely resembles real audio. More notably, when processing lossy MP3 input, baseline models exhibit significant performance degradation due to the lack of compression artifact recovery capability, while our Vocal VAE shows only modest performance decline, demonstrating robustness to lossy inputs. The high-quality acoustic reconstruction establishes a solid upper bound for subsequent singing voice generation tasks.

\textbf{Performance on Multi-Singer Generation.} Table \ref{tab:main_results} presents the comparison results of all models on objective and subjective metrics. It should be noted that Tutti is the first generation framework supporting dynamic multi-singer arrangement within a single song, while existing baselines such as Vevo2 can only handle single-singer scenarios, thus preventing direct comparison on multi-singer related metrics. On comparable basic metrics, Tutti demonstrates significant advantages in lyric intelligibility, with WER notably lower than Vevo2, benefiting from our word-level lyric alignment strategy. Regarding timbre consistency, since we employ a duration-weighted averaging strategy for calculating similarity in multi-singer scenarios, Tutti achieves higher SIM than Vevo2, indicating that the structure-aware singer prompt can maintain stable identity representations in complex singer switching scenarios. In subjective evaluations, Tutti outperforms Vevo2 in both audio quality and naturalness. Notably, Vevo2 occasionally exhibits interruption issues in longer audio, whereas Tutti demonstrates more robust fluency performance owing to its DiT-based long-sequence modeling capability. For the multi-singer singing task, we introduce two specialized metrics: MS-MOS and Mel-MOS. Regarding multi-singer distinctiveness, Tutti achieves a high MS-MOS score, indicating that individual singer timbres can be clearly distinguished in multi-singer scenarios while maintaining good fusion and balance between vocal parts. For melody performance, although Mel-MOS still shows a gap compared to Ground Truth, the overall melodicity and pitch accuracy have reached a relatively high level.

\begin{figure}[tb!]
	\centering
	\makebox[\columnwidth][c]{\includegraphics[width=1.1\columnwidth]{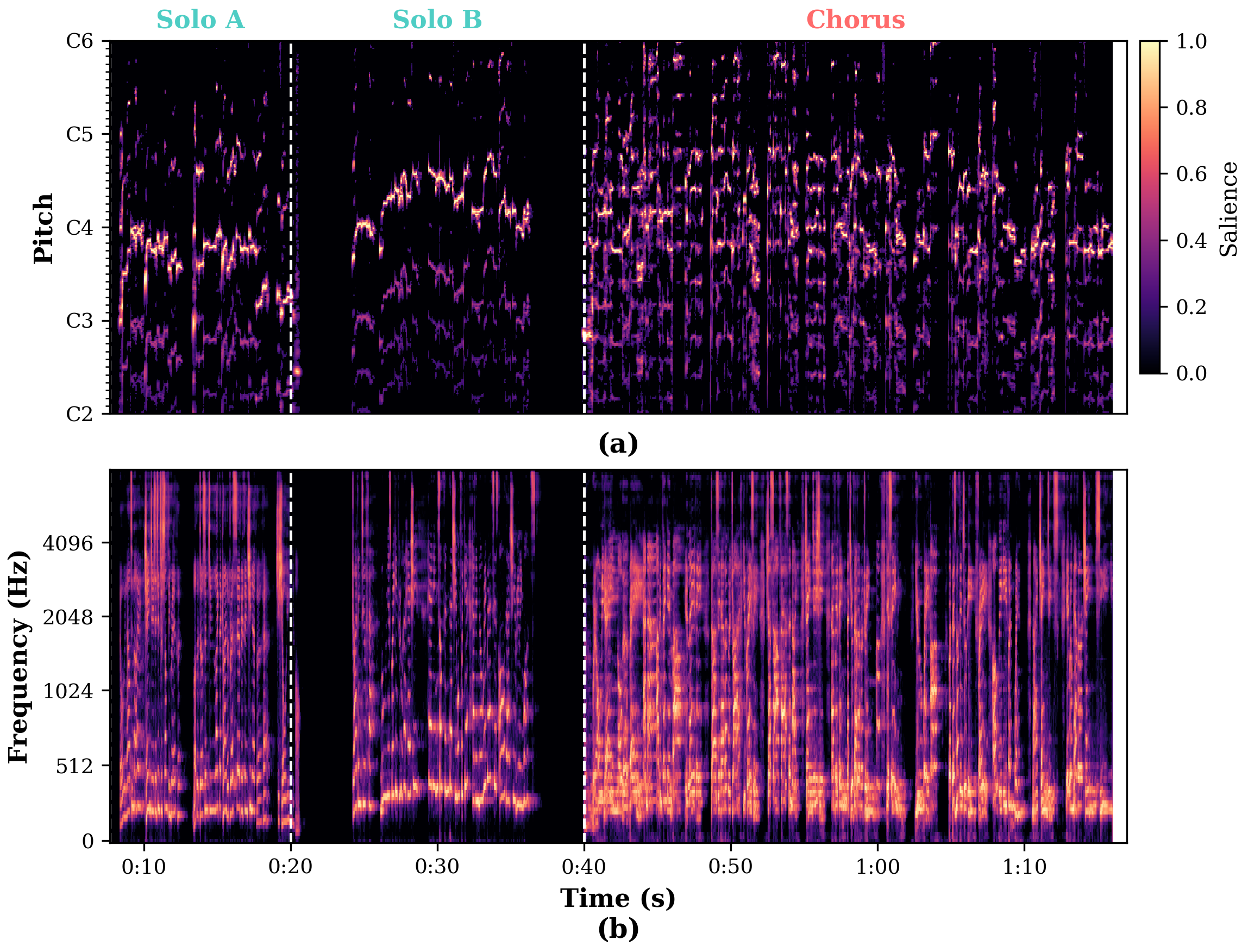}}
    \caption{Acoustic visualization analysis of generated multi-singer audio. (a) The Pitch Salience Map depicts distinct melodic patterns, showing single trajectories for solos and interwoven lines for the chorus. (b) The Mel Spectrogram illustrates the energy distribution in the time-frequency domain, highlighting spectral coherence and rich resonance structures.}
    \label{fig:chorus_analysis}
\end{figure}

To intuitively verify the capability of our model in multi-singer scheduling and choral modeling, we conducted an acoustic visualization analysis of the generated audio. As shown in Figure \ref{fig:chorus_analysis}, we present the Pitch Salience Map and Mel Spectrogram containing both solo and choral segments. (a) In the Pitch Salience Map, white dashed lines clearly mark the structural boundaries between Solo A (0:00-0:20), Solo B (0:20-0:40), and Chorus (0:40 onwards), indicating that the model precisely adheres to the input structural controls. In the Solo A and Solo B regions, the pitch map exhibits clear, single fundamental frequency trajectories primarily concentrated in the C3-C4 range, characteristic of individual vocalists. In contrast, the Chorus region displays a markedly different pattern: the pitch energy distribution significantly widens to span C3-C5, with multiple interweaved melodic contours visible simultaneously. This distinct harmonic layering phenomenon—where parallel pitch trajectories appear at different frequency levels—directly reflects the multi-part harmonic characteristics of choral singing. (b) The Mel Spectrogram provides complementary evidence from the spectral perspective. The Solo regions show relatively sparse energy distribution with clear formant structures, while the Chorus region exhibits substantially higher energy density across the entire frequency range, richer high-frequency harmonic content, and more continuous temporal energy patterns. This is consistent with the acoustic characteristic of sound pressure superposition in multi-person singing, further demonstrating that Tutti successfully models the complex acoustic texture and interaction in choruses.

\textbf{Effect of Adaptive Singer Prompt Fuser.} Removing this component leads to significant degradation on the multi-singer distinctiveness metric, accompanied by notably increased score variance. This indicates that simple linear weighted summation fails to handle the complex primary-secondary relationships between lead vocals and harmonies in choral scenarios. Without the non-linear interaction modeling provided by the self-attention mechanism, the timbre boundaries between parts become blurred, resulting in unstable singer distinctiveness in generated choral segments. Furthermore, this variant also exhibits degradation in lyric intelligibility, with some syllables becoming ambiguous, likely due to the lack of an effective singer scheduling mechanism leading to insufficient coupling between timbre representation and content generation. Additionally, qualitative analysis reveals that in scenarios requiring extremely rapid singer transitions (\textit{e.g.}, $<$ 0.2s), the absence of this module leads to severe timbre tailing, where the preceding singer's timbre unnaturally persists into the subsequent segment.

\textbf{Effect of Complementary Texture Learning.} The ablation of this module results in a notable decline on the melody performance metric, with significantly increased score variance. This phenomenon validates the importance of texture modeling for capturing acoustic details in choruses. Without the guidance of texture features, the generated audio lacks acoustic qualities such as spatial reverberation and spectral fusion, leading to reduced melodic richness and realism. Notably, this variant also exhibits some degradation in audio quality metrics, suggesting that texture modeling contributes positively not only to the overall choral atmosphere but also to frame-level acoustic quality. Interestingly, this variant shows slightly improved timbre consistency, which we hypothesize is because removing texture modeling forces the model to rely more heavily on explicit singer conditions for generation, thus exhibiting stronger condition adherence in the timbre dimension, albeit at the cost of sacrificing acoustic texture richness.

\section{Limitations and Future Work}

We acknowledge the limitations of our proposed framework. First, our method relies on the strong assumption that verse sections contain only a single singer. However, in real-world musical compositions, verses may involve multiple singers or duets. This simplified assumption limits the model's ability to handle such complex arrangement scenarios within verse sections. Second, regarding the representation of singing voices, we find that the adopted features do not fully cover the vocal characteristics of mainstream singers, which limits our ability to utilize certain specific timbres for performance. Third, we observe that there is still significant room for improvement in the melodicity of the generated singing voice, as it currently falls short of delivering truly compelling and emotionally resonant performances. Therefore, our future work will focus on exploring more efficient compressed representations of singing voice features and striving to restore more infectious and expressive singing performances.

\section{Conclusion}

In this paper, we introduced Tutti, the first generative framework designed for dynamic multi-singer arrangement within a single song, aiming to transcend the limitations of the soloist paradigm in traditional singing voice synthesis. By introducing a Structure-Aware Singer Prompt, we achieved dynamic role scheduling that evolves with the musical structure, enabling the model to flexibly handle complex transitions between solos and choruses. Furthermore, we proposed Complementary Texture Learning via Condition-Guided VAE, which successfully captures the complex acoustic textures within multi-part interactions utilizing a condition-guided reconstruction task. Experimental results demonstrate that Tutti not only excels in the precision of multi-singer arrangement but also significantly enhances the acoustic realism and artistic cohesion of choral generation, paving a new path for complex multi-singer arrangement.


\newpage
\bibliography{example_paper}
\bibliographystyle{icml2026}

\newpage
\appendix
\onecolumn
\section{Implementation and Training Details}

\subsection{Model Configuration}
For the Vocal VAE, we adopt the architecture of Stable Audio 2.0 \cite{evans2025stable}, operating at a sampling rate of 44.1kHz with a latent dimension of 64, consistent with the original configuration. The DiT backbone is implemented as a 16-layer Transformer with a hidden dimension of 2048, 32 attention heads, and 4-layer Feed-Forward networks. Regarding the condition embeddings, the dimensions are set as follows: lyrics embedding at 512, structure label embedding at 20, singer prompt embedding at 192, texture embedding at 64, and time embedding at 512. The mel-spectrogram latent dimension matches the VAE latent size of 64.

\subsection{Training Settings}
All models are trained using 16 NVIDIA A100 GPUs. We consistently employ the AdamW optimizer \cite{loshchilov2019adamw} across all training stages.

\textbf{VAE Training.} For the Vocal VAE, the optimizer hyperparameters are set to $\beta_1=0.8$ and $\beta_2=0.99$. The model is trained with a batch size of 32 and an initial learning rate of $1.5 \times 10^{-4}$ for a total of 970K steps. For the Condition-Guided VAE training phase (Texture Learning), we adjust the batch size to 8 and reduce the learning rate to $5 \times 10^{-5}$, training for an additional 200K steps. To facilitate robust texture feature extraction, Gaussian noise perturbation is applied to the latent space with a Signal-to-Noise Ratio (SNR) of approximately 10dB during this phase.

\textbf{DiT Backbone Training.} The DiT model is trained with a batch size of 4 and a learning rate of $1 \times 10^{-5}$. We utilize a learning rate scheduler combining linear warmup and linear decay. The training process spans 950K steps.

\subsection{Inference Configuration}
During inference, we employ the Euler ODE solver to generate samples. The sampling process is performed with 50 inference steps. To enhance generation quality and condition adherence, we apply Classifier-Free Guidance (CFG) with a guidance scale of 4.0.

\section{Subjective Evaluation Details}
\label{app:subjective}

To ensure a comprehensive assessment of the generated singing voices, we organized a subjective evaluation involving 30 participants, divided into a professional group (10 experts) and a general group (20 volunteers). We followed the standard Mean Opinion Score (MOS) protocol, where participants were asked to rate the samples on a 5-point Likert scale (1: Bad, 2: Poor, 3: Fair, 4: Good, 5: Excellent) with 0.5 increments.

The professional group was recruited from music conservatories and professional art troupes. Given that vocal performance requires extensive physiological training, we imposed strict selection criteria regarding their domain expertise. As shown in Table \ref{tab:evaluators}, the group covers three major singing styles: Bel Canto, Chinese Folk, and Pop/Jazz, along with experts in Choral Conducting to specifically evaluate multi-singer harmony. The average training duration of these experts is 13.7 years. The general group consists of 20 volunteers with diverse backgrounds and normal hearing, representing the perspective of ordinary listeners. They were asked to focus on the overall naturalness and melody of the songs.

\begin{table}[h]
\caption{Demographic profiles of the 10 professional evaluators.}
\label{tab:evaluators}
\begin{center}
\resizebox{0.8\columnwidth}{!}{
\begin{tabular}{c|l|l|c|l}
\toprule
ID & Major/Profession & Academic Status & Training (Yrs) & Area of Expertise \\
\midrule
P01 & Bel Canto (Soprano) & PhD Candidate & 18 & Opera performance, Resonance control \\
P02 & Bel Canto (Tenor) & Professional & 15 & Vocal technique, Stage performance \\
P03 & Chinese Folk Vocal & Master Student & 12 & Folk style, Diction, Regional characteristics \\
P04 & Pop Vocal Performance & Instructor & 20 & Pop style, Microphone technique, Expression \\
P05 & Jazz Vocal & Senior Undergrad & 10 & Improvisation, Rhythm, Jazz harmony \\
P06 & Choral Conducting & Professional & 16 & Choral balance, Part blending, Rehearsal \\
P07 & Choral Conducting & Master Student & 13 & Intonation, Multi-part coordination \\
P08 & Vocal Composition & PhD Candidate & 15 & Vocal writing, Counterpoint, Harmony \\
P09 & Vocal Pedagogy & Master Student & 11 & Voice physiology, Training methods \\
P10 & Music Production (Vocal) & Freelance & 9 & Vocal mixing, Texture processing \\
\bottomrule
\end{tabular}
}
\end{center}
\end{table}

\section{Verification of Non-Leakage in Texture Learning}
\label{app:texture_leakage}

To verify that the Condition-Guided VAE learns texture representations that are independent of melodic content, lyrical information, and speaker identity, we conducted a systematic texture swapping experiment.

\subsection{Experimental Setup}

We randomly selected 50 generated samples for evaluation. For each sample, we fixed the random seed to 42 and established a baseline by generating audio without any texture reference (denoted as ``No Texture''). We then introduced 10 distinct reference audio clips as texture inputs, which were randomly selected to cover a wide range of acoustic scenarios, including solo performances, duets, and multi-singer choruses, with diverse melodies and lyrics. This setup results in a total of 500 generated variations. We compare these textured generations against the baseline to evaluate melodic consistency, and calculate standard objective metrics to assess content and identity stability.

\subsection{Results and Analysis}

\begin{table}[h]
\caption{Verification of non-leakage in texture learning. We report the average metrics across all 500 generated samples using 10 different texture references. F0 metrics measure the deviation from the ``No Texture'' baseline, while WER and SIM evaluate the absolute performance.}
\label{tab:texture_leakage}
\begin{center}
\resizebox{0.4\columnwidth}{!}{
\begin{tabular}{lc}
\toprule
\textbf{Metric} & \textbf{Value} \\
\midrule
\multicolumn{2}{l}{\textit{Melodic Consistency (vs. No Texture)}} \\
F0 RMSE (cents) $\downarrow$ & 7.85 \\
F0 MAE (cents) $\downarrow$ & 6.21 \\
Correlation $\uparrow$ & 0.9997 \\
\midrule
\multicolumn{2}{l}{\textit{Content \& Identity (Absolute)}} \\
WER $\downarrow$ & 13.53\% \\
SIM $\uparrow$ & 0.690 \\
\bottomrule
\end{tabular}}
\end{center}
\end{table}

As presented in Table \ref{tab:texture_leakage}, the introduction of diverse texture references has a negligible impact on the explicit attributes of the generated audio. The F0 contours of the textured generations remain tightly aligned with the non-textured baseline, with an RMSE of less than 8 cents and a near-perfect correlation of 0.9997, demonstrating that the texture encoder does not leak melodic information. Furthermore, the Word Error Rate (13.53\%) and Speaker Similarity (0.690) remain comparable to the main results reported in Table \ref{tab:main_results} (13.50\% and 0.691, respectively), confirming that the texture representation captures complementary acoustic properties without interfering with the linguistic content or singer identity defined by the explicit conditions.

confirming that the texture representation captures complementary acoustic properties without interfering with the linguistic content or singer identity defined by the explicit conditions.

\section{Analysis of Adaptive Singer Prompt Fuser Mechanism}
\label{app:fuser_analysis}

To investigate how the Adaptive Singer Prompt Fuser handles multi-singer scenarios and verify that it effectively aggregates multi-singer information without losing individual characteristics, we visualize the attention weights within the fuser module.

\begin{figure}[h]
    \centering
    \includegraphics[width=0.9\columnwidth]{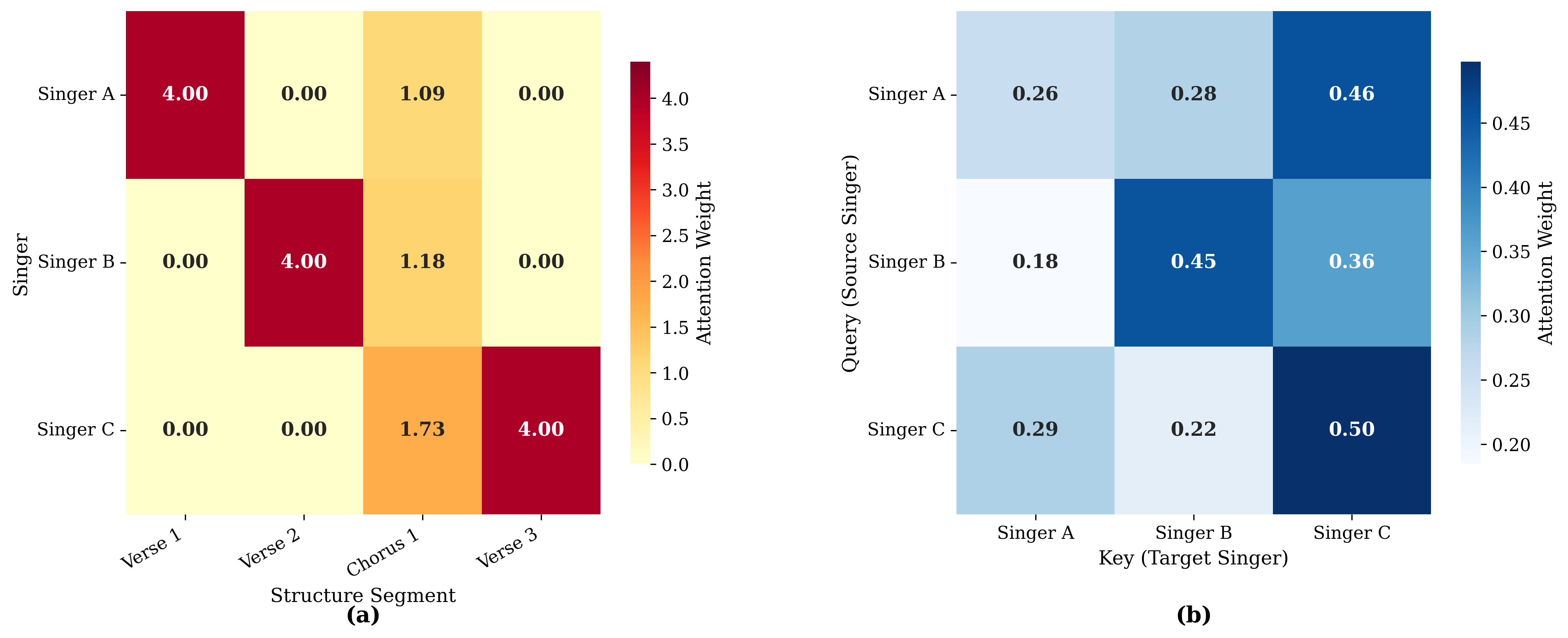}
    \caption{Visualization of attention weights in the Adaptive Singer Prompt Fuser. (a) The Singer Attention Matrix shows the weight distribution of candidate singers across different structural segments. (b) The Singer-to-Singer Cross-Attention Matrix details the internal interaction weights within the multi-singer Chorus 1 segment.}
    \label{fig:fuser_analysis}
\end{figure}

Figure \ref{fig:fuser_analysis} presents the visualization of the attention mechanism. Figure \ref{fig:fuser_analysis}(a) illustrates the attention weights assigned to three candidate singers (Singer A, B, C) across four structural segments. In the solo sections (Verse 1, Verse 2, and Verse 3), the fuser exhibits a highly selective attention pattern, assigning a dominant weight (4.00) to the specific soloist while suppressing others (0.00). This confirms the model's capability for precise singer scheduling. Crucially, in the multi-singer section (Chorus 1), the attention is distributed among all three singers (1.09, 1.18, 1.73). This demonstrates that the resulting singer embedding is not a collapse to a single identity but a weighted composite that retains the acoustic signatures of all participants.

Furthermore, Figure \ref{fig:fuser_analysis}(b) details the internal self-attention mechanism within the Chorus 1 segment, revealing the interaction logic between singers. The matrix shows non-uniform attention weights, indicating that the fusion is not a simple average. For instance, Singer A (as Query) attends significantly to Singer C (0.46) and Singer B (0.28) rather than solely to itself. This suggests that the fuser models the harmonic relationships and timbral blending between singers, effectively encoding the ``ensemble texture'' into the single embedding. This mechanism ensures that even when compressed into a single vector, the representation encapsulates the complex interplay of multiple voices, allowing the subsequent generation model to decode and reconstruct the multi-singer performance accurately.

\section{Data Analysis on Singer Consistency in Verse Sections}
\label{app:verse_analysis}

\begin{figure}[h]
    \centering
    \includegraphics[width=0.6\columnwidth]{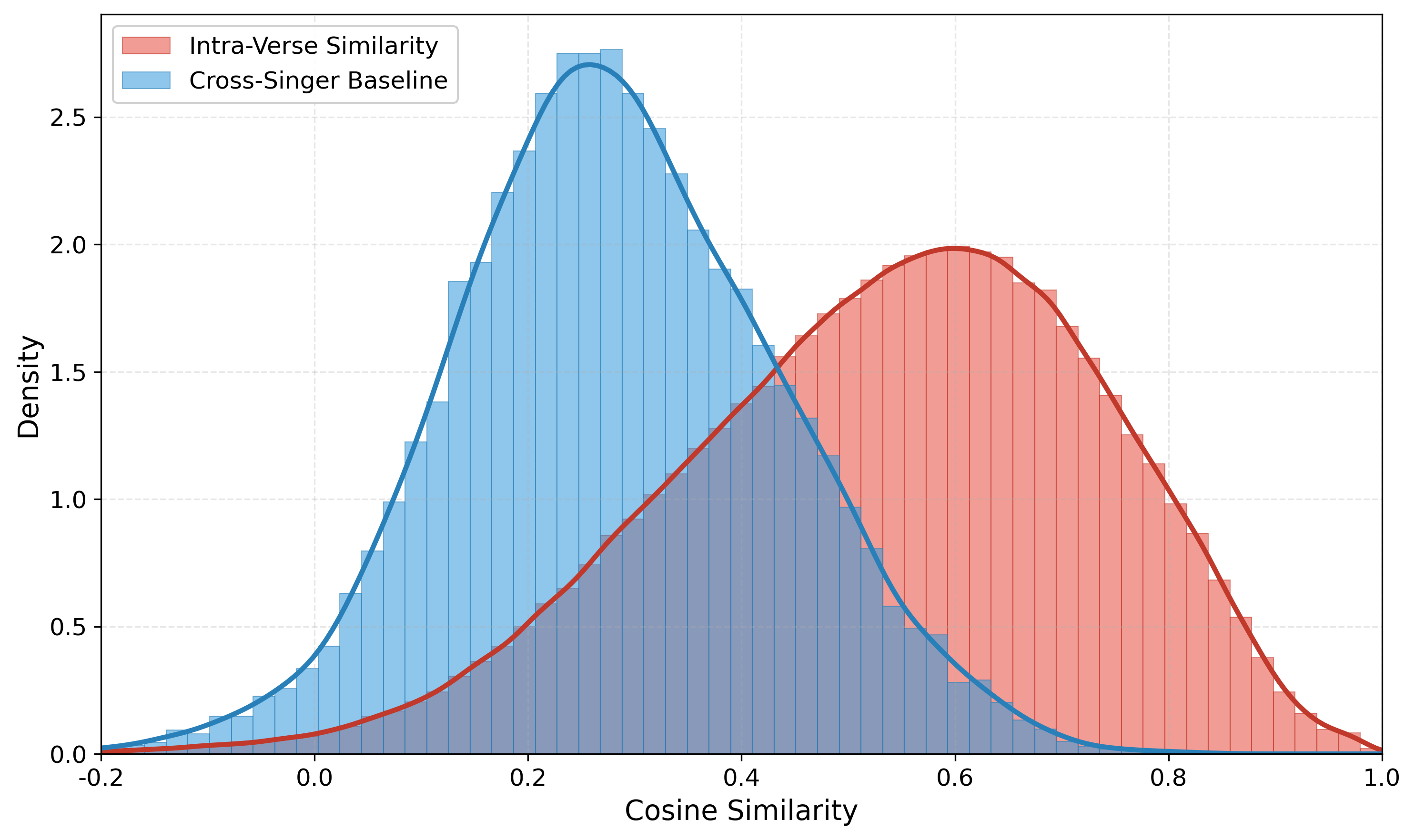}
    \caption{Distribution of cosine similarity for Intra-Verse segments (red) and Cross-Singer baseline (blue). The significant separation between the two distributions supports the assumption that verse sections are predominantly sung by a single vocalist.}
    \label{fig:verse_analysis}
\end{figure}

To empirically validate the assumption that verse sections typically feature a single vocalist (Section \ref{sec:structure-aware singer prompt}), we conducted a large-scale statistical analysis on the singer consistency within verse segments.

\subsection{Experimental Setup}
We randomly sampled approximately 10,000 songs from our dataset, ensuring each sample contained valid structure labels and passed voice activity detection. We designed two comparative groups to analyze the distribution of singer similarity:

\textit{Intra-Verse Similarity:} For each song, we extracted multiple non-silent segments (approx. 3 seconds each with 1.5-second overlap) from the same verse section. We then computed the cosine similarity between the speaker embeddings of these segments, resulting in a total of 323,996 pairs. This group represents the singer consistency within a single verse.

\textit{Cross-Singer Baseline:} We randomly paired segments from different songs to simulate the similarity distribution between different singers. This group consists of 100,000 pairs and serves as a baseline for non-identity matches.

\subsection{Statistical Evidence and Discussion}

Figure \ref{fig:verse_analysis} illustrates the probability density functions of the cosine similarities for both groups. The quantitative results reveal a distinct separation between the two distributions. Specifically, the Intra-Verse pairs exhibit a significantly higher similarity, with a mean of 0.5355 and a median of 0.5526 (std=0.1973). In contrast, the Cross-Singer baseline shows a much lower distribution, with a mean of 0.2835 and a median of 0.2778 (std=0.1515).

While the absolute similarity scores for the same singer in the singing domain (around 0.54) are lower than those typically observed in speech tasks (often $ >0.7 $)—a phenomenon attributed to the acoustic variability of singing (e.g., pitch shifts, phonation changes) and the domain gap of the speech-pretrained CAM++ model—the relative margin is substantial. The significant distributional gap between the intra-verse and cross-singer pairs empirically supports our assumption that verse sections are predominantly performed by a single vocalist. Furthermore, this distribution justifies our threshold selection in the data processing pipeline (e.g., 0.4), which effectively separates the majority of same-singer segments from different-singer pairs.


\end{document}